\documentclass{ws-procs975x65}

\begin{document}
\hfill{\parbox[b]{1in}{ \hbox{\tt PNUTP-13/A01} }}
%\vskip -0.2in
\title{Composite Higgs and Techni-Dilaton at LHC}

\author{Deog Ki Hong$^*$}

\address{Department of Physics, Pusan National University,\\
Busan, 609-735, Korea\\
$^*$E-mail: dkhong@pusan.ac.kr}

\begin{abstract}
In this talk I argue that a light techni-dilaton exists in models of walking technicolor near the conformal window.
The light techni-dilaton then mixes with the composite Higgs, which is predicted to be narrow and light in walking technicolor models. 
It is an interesting possibility that the newly discovered boson at LHC might be a mixed state of techni-dilaton and the composite Higgs.  
\end{abstract}


\bodymatter

\section{Introduction and Review}\label{aba:sec1}
\begin{figure}[tbh]
	\centering
	\includegraphics[angle=0,width=0.44\textwidth]{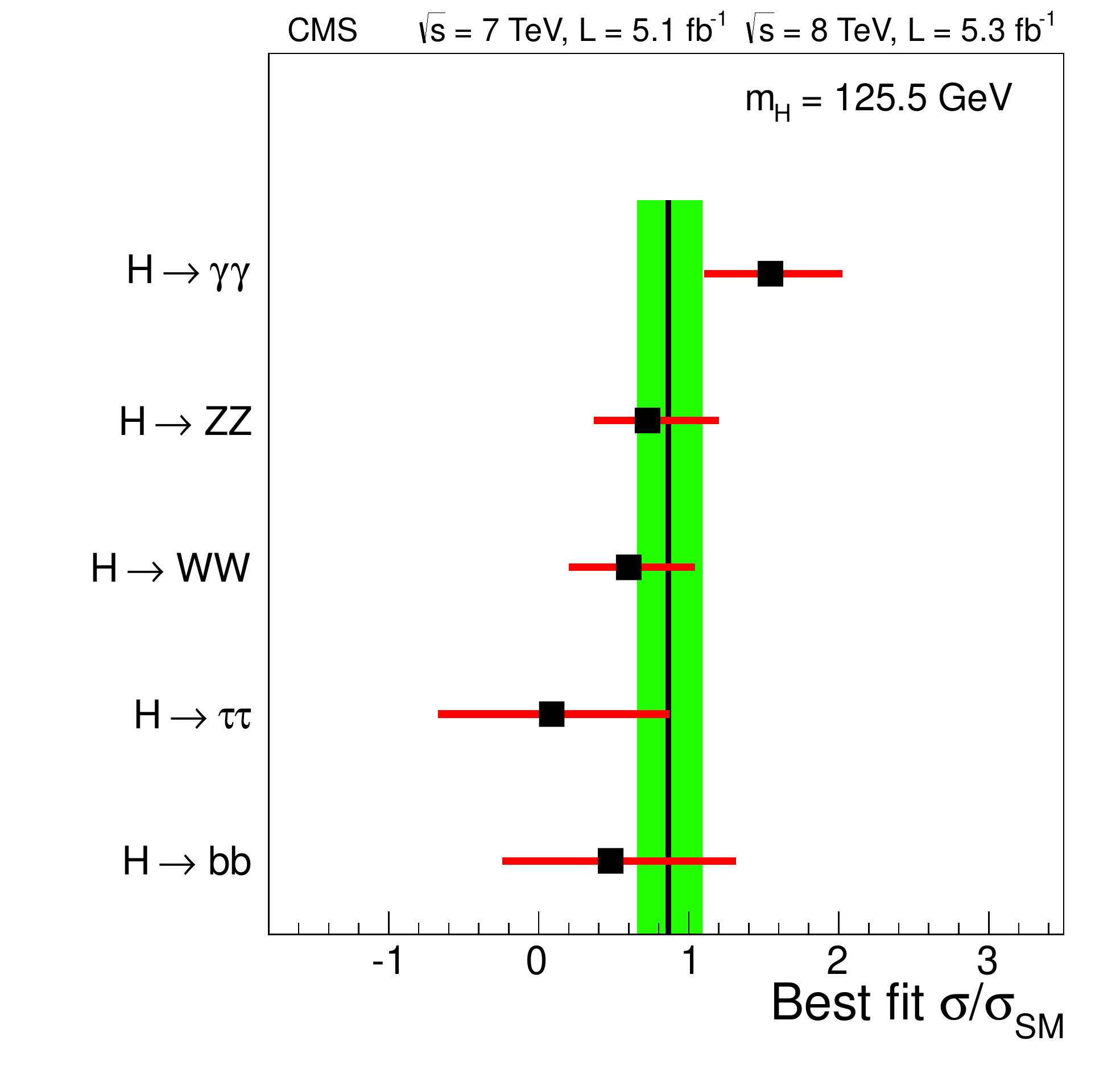}
\hskip 0.1in
	\includegraphics[angle=0,width=0.52\textwidth]{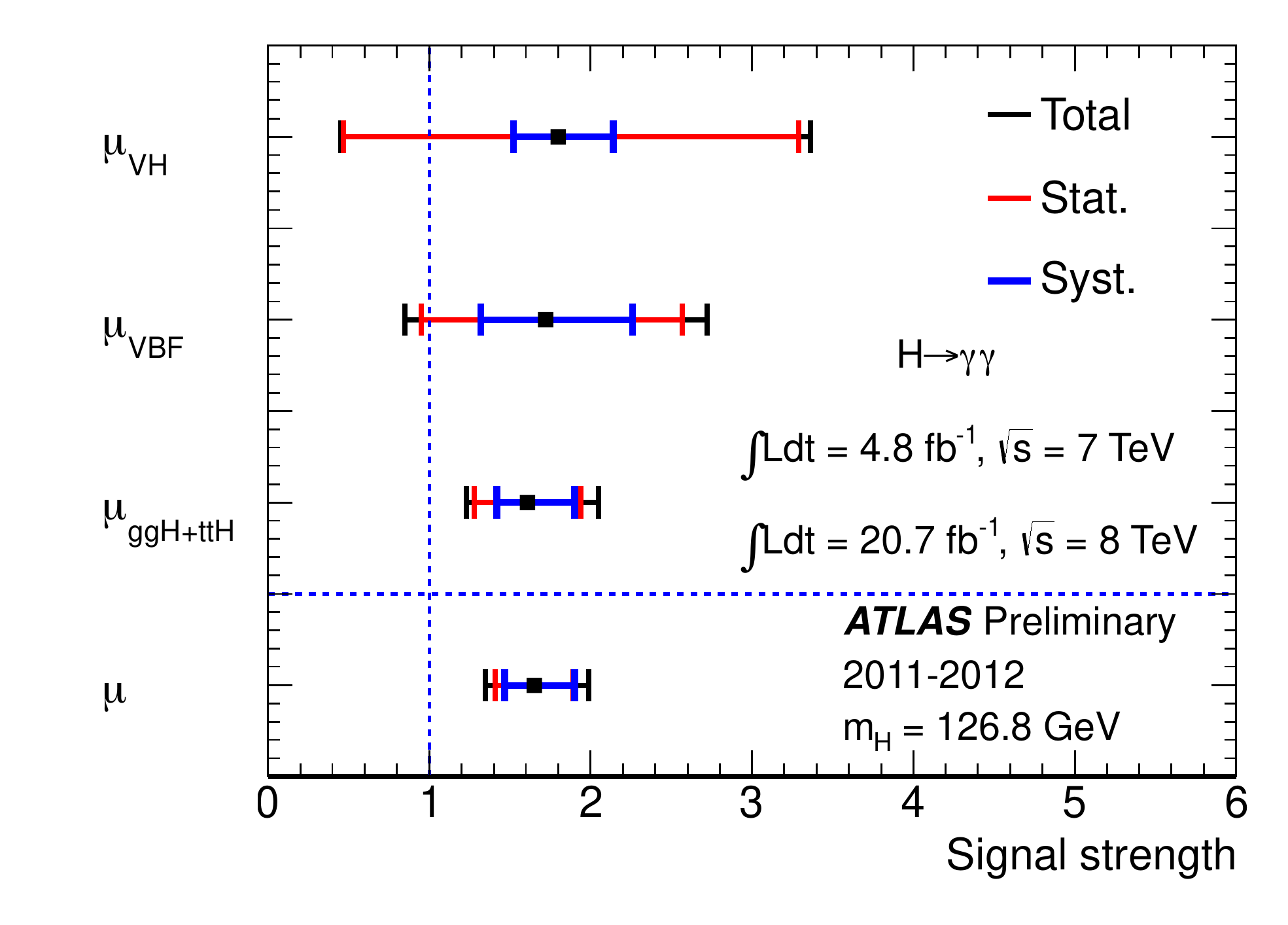}
\label{fig1}
\caption{The CMS results on the ratio of the cross sections of Higgs decay channels over the SM values (left panel) and the $\mu$ values of Higgs to two-photon channels, measured at ATLAS (right panel).}
\end{figure}

A new boson of mass $125\,{\rm GeV}$ has been discovered at the Large Hadron Collider (LHC)~\cite{:2012gu,:2012gk}.
Since the discovery was made, more data on the new boson has been accumulated to show that it is very much like the standard model Higgs~\cite{moriond}. The data is now consistent  more or less with the standard model (SM) of elementary particles with deviation less than $2\sigma$ (See Fig.\ref{fig1}). % and the statistical error will be reduced significantly at LHC14. 

But, there are several convincing arguments that the SM is not complete and there should be new physics beyond the standard model (BSM).  For instance, it does not explain the origin of dark matter, which constitutes 27\% of total energy of our 
universe, according to the recent analysis of Planck 2013 data on the cosmic microwave background~\cite{Ade:2013skr}. 
It is also known that the SM vacuum becomes meta-stable at energy $10^{10}$-$10^{11}~{\rm GeV}$ if we assume the SM is valid up to the Planck scale~\cite{Degrassi:2012ry}. 
So, the fact that we have not  seen any significant deviation of the SM and no new BSM  particles at the LHC may just mean that 
the scale of new physics is much higher than the scale, probed at the LHC 
with the center of mass energy $\sqrt{s}=8~{\rm TeV}$.

In this talk I try to argue that the small deviation, though less than $2\sigma$, seen at the LHC8 might be due to walking technicolor (WTC), a model of strong dynamics BSM, which breaks the electroweak symmetry dynamically by a new strong force.  
Especially, the newly discovered boson might be a mixture of two light scalars, namely light techni-dilaton and light composite Higgs, that WTC  predicts.

\section{Light Dilaton and PCDC}
\begin{figure}[tbh]
%\vskip 0.15in
	\centering
	\includegraphics[width=0.4\textwidth]{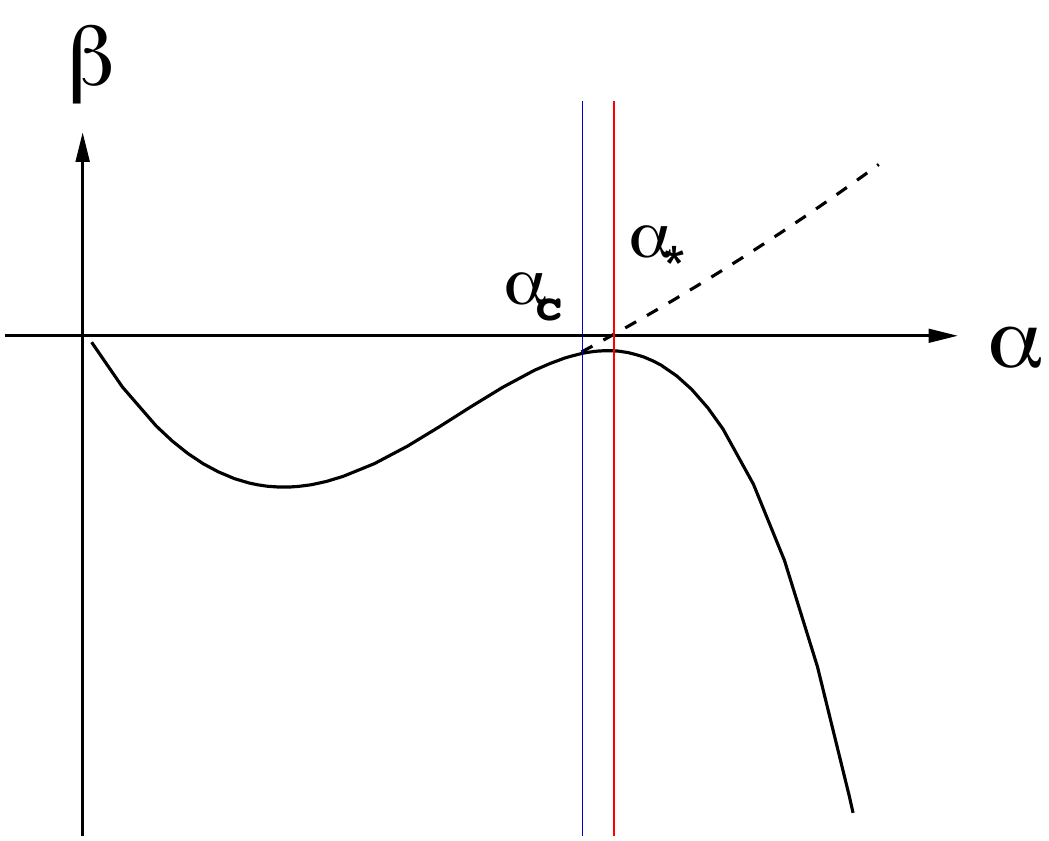}
	\hskip 0.1in \includegraphics[width=0.5\textwidth]{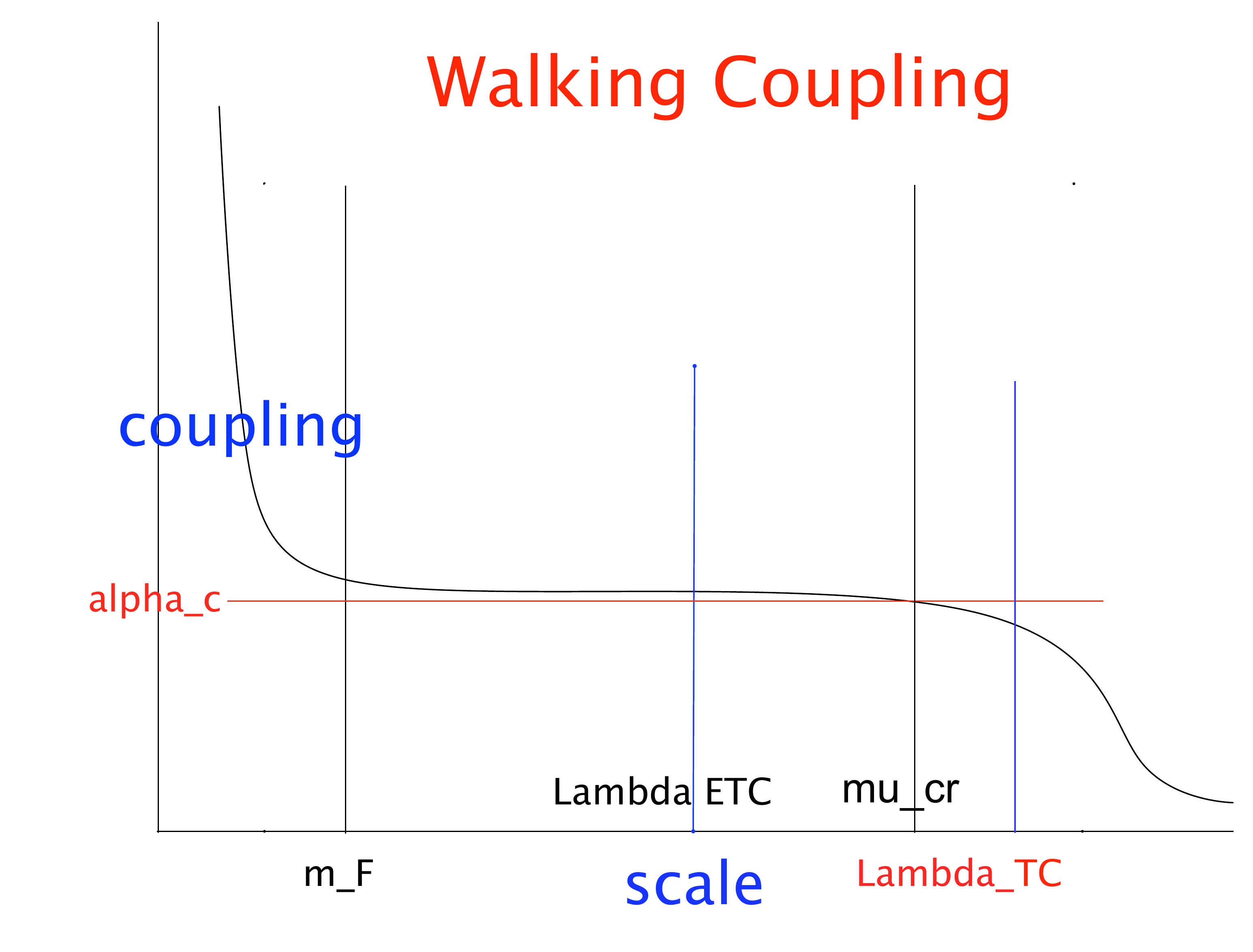}
	\caption{The $\beta$ function of WTC is shown in the left panel and the scale dependence of the coupling is plotted in the right panel.}
	\label{fig2}
\end{figure}

Walking technicolor is a model for dynamical electroweak symmetry breaking, where the electroweak symmetry is broken by the 
condensation of techni-quarks, with a nontrivial quasi infra-red (IR) fixed point~\cite{Holdom:1981rm,Yamawaki:1985zg,Akiba:1985rr}. The beta function of WTC and the scale dependence of the technicolor coupling are shown in Fig.~\ref{fig2}. 

If the condensation of techni-quarks were not to occur, the coupling would have an IR fixed point,  $\beta(\alpha_*)=0$,  
and theory will flow into a conformal window (CW). However, since in WTC we need to generate a scale at IR to break the electroweak symmetry, the theory should be slightly away from the conformal window so that the condensation of techni-quarks does occur. Namely the critical coupling for chiral symmetry breaking should be smaller than $\alpha_*$  and should be reached at a scale $\mu_{\rm cr}$ before the theory flows into the IR fixed point. If the critical coupling is very close to the would-be IR fixed point, $\alpha_c\equiv\alpha(\mu_{\rm cr})\approx\alpha_*$,  the $\beta$ function almost vanishes 
and thus the coupling remains nearly constant for scales between $\mu_{\rm cr}$ and $m_F$, the dynamical mass of techni-quarks, which is of order of $1~{\rm TeV}$. 
Below the scale $m_F$ the techni-quarks decouple and the coupling gets very strong to confine the techni-gluons. 

The Schwinger-Dyson analysis for the techni-quark bilinear operator shows that its anomalous dimension is very close to 1
for $m_F<\mu\le\mu_{\rm cr}$,
\begin{equation}
\gamma_m(\mu)=1+\sqrt{\alpha(\mu)/\alpha_c-1}\approx1\,.
\end{equation}
Therefore, in WTC the techni-quark condensation is quite enhanced in the ultra-violet (UV) region or 
at the extended technicolor (ETC) scale, $\Lambda_{\rm ETC}<\mu_{\rm cr}$ so that 
ETC interactions give large masses to SM fermions, while suppressing the contributions of ETC interactions to the flavor-changing neutral currents by raising the ETC scale high enough, $\Lambda_{\rm ETC}\gtrsim 10^3\,m_F$:
\begin{equation}
\left. \left<\bar QQ\right>\right|_{\Lambda_{\rm ETC}}=e^{\int_{m_F}^{\Lambda_{\rm ETC}}
\frac{{\rm d}\mu}{\mu}\gamma_m(\mu)}\left. \left<\bar QQ\right>\right|_{m_F}
=\frac{\Lambda_{\rm ETC}}{m_F}\left.\left<\bar QQ\right>\right|_{m_F}\approx m_F^2\,\Lambda_{\rm ETC}\,.
\end{equation}

\begin{figure}[tbh]
%\vskip 0.15in
	\centering
	\includegraphics[width=0.4\textwidth]{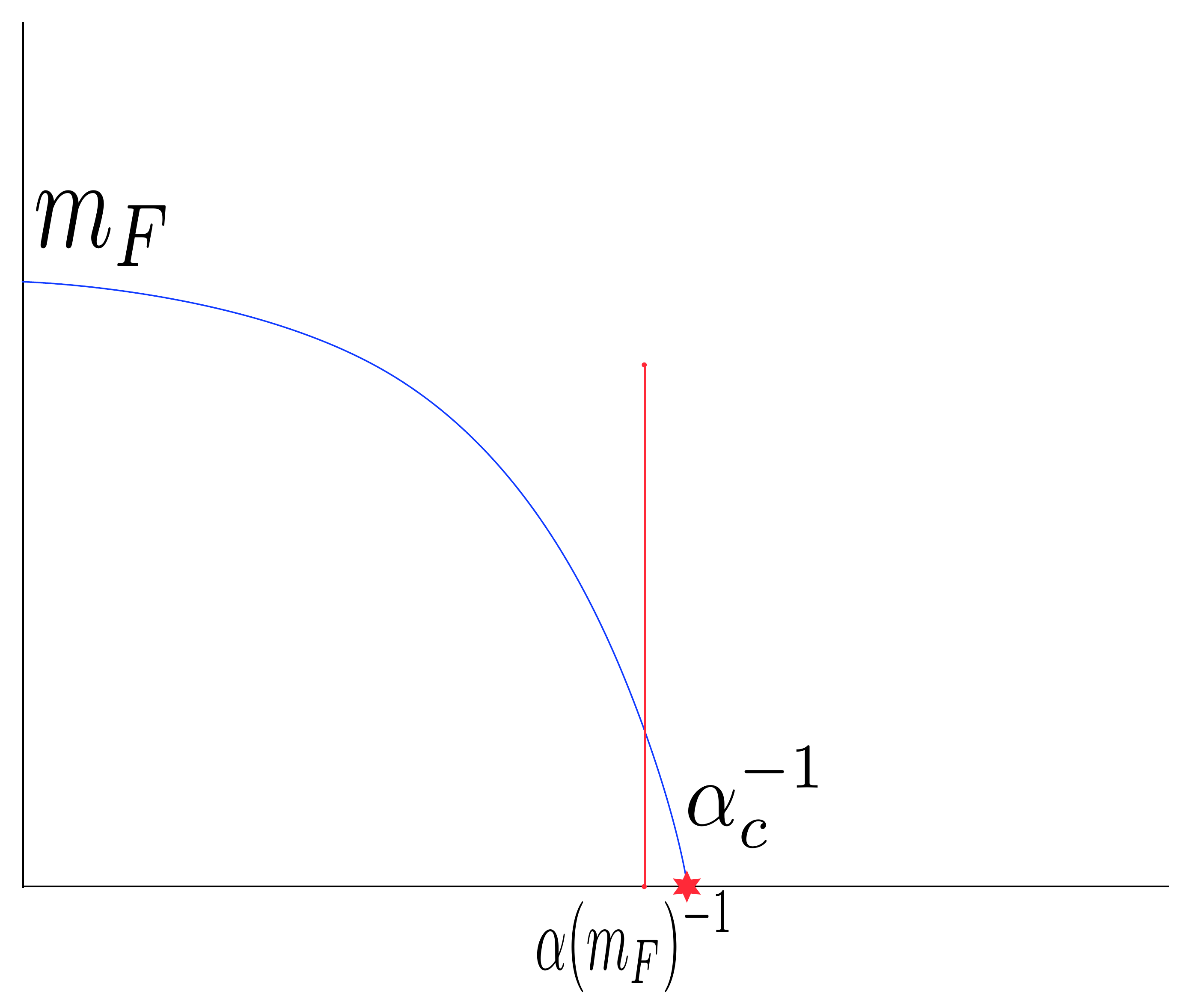}
	\caption{The quantum (conformal) phase transition in a theory space as one varies $\alpha(m_F)$.}
	\label{mir}
\end{figure}

In a theory space, one can approach the conformal window by tuning the number of flavors. For ${\rm SU}(N_{\rm TC})$ gauge theories with $N_f$ fermions in the fundamental representation, the two-loop calculation of $\beta$ function shows that theories approach a conformal window for $x=N_f/N_{\rm TC}\approx4$ in the large $N_{\rm TC}$ limit~\cite{Appelquist:1996dq}. 
Another, but equivalent way of probing the theory space in the chirally broken phase is to change the coupling at the scale of dynamical mass, $\alpha(m_F)$. Miransky and Yamawaki has shown that the quantum phase transition that occurs, when $\alpha(m_F)\to\alpha_c$, exhibits an essential singularity, so-called Berezinsky-Kosterlitz-Thouless (BKT) or Miransky scaling (see Fig.~\ref{mir})~\cite{Miransky:1996pd}. Namely, the order parameter for the phase transition behaves near the phase transition as 
 \begin{equation}
%\hskip -1 in 
m_F=\Lambda_{\rm TC}\,\exp\left(-\frac{\pi}{\sqrt{\frac{\alpha(m_F)}{\alpha_c}-1}}\right)\,,
\end{equation}
where $\Lambda_{\rm TC}$ is the intrinsic scale of WTC, shown in Fig.~\ref{fig2}.

Since  the coupling is almost constant and $\beta(\alpha)\approx0$, 
WTC has an approximate scale invariance, broken spontaneously, for scales $m_F <\mu<\mu_{\rm cr}$.
Therefore there exists a dilatation current, $D_{\mu}=x^{\nu}\theta_{\mu\nu}$, where $\theta_{\mu\nu}$ is 
the so-called improved energy-momentum tensor. The dilatation current is, however, anomalous due to quantum effects. 
By an explicit calculation in the ladder approximation the scale anomaly is found to be proportional to $m_F^4$:
\begin{equation}
\left<\theta^{\mu}_{\mu}\right>=\frac{\pi\beta({\alpha})}{\alpha^2}\left<\left(F_{\mu\nu}^a\right)^2\right>+\frac{g^2}{\mu_{\rm cr}^2}\left<\left(\bar QQ\right)^2\right>\sim m_F^4.
\end{equation}
\begin{figure}[tbh]
%\vskip 0.15in
	\centering
	\includegraphics[width=0.7\textwidth]{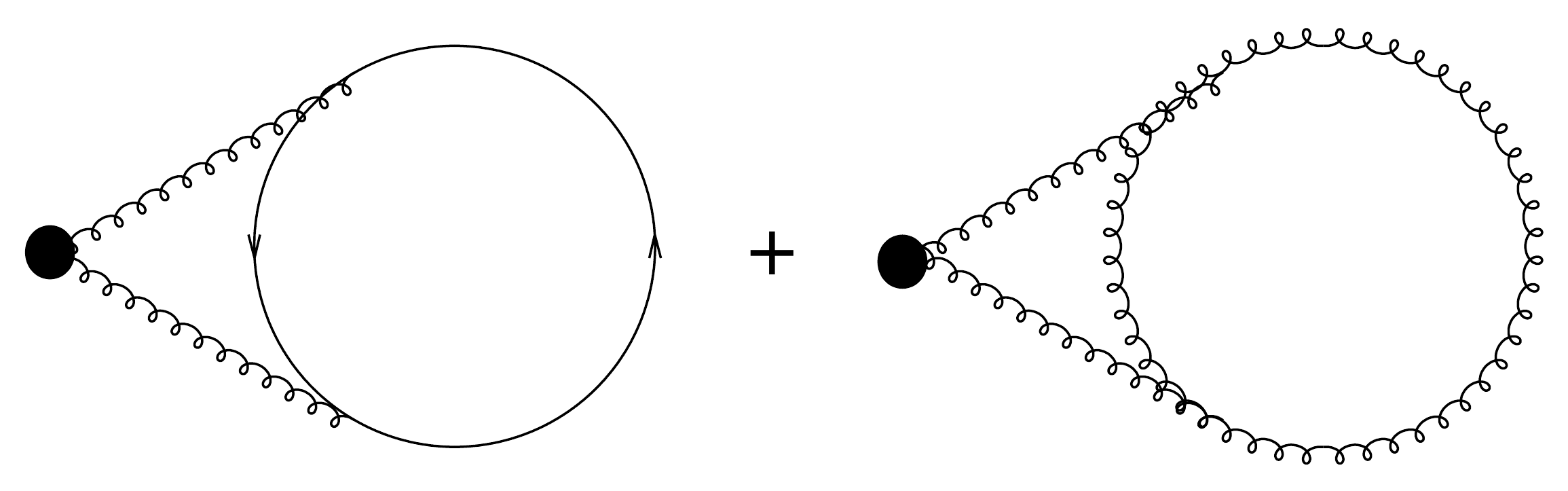}
	\hskip 0.1in
	 \includegraphics[width=0.2\textwidth]{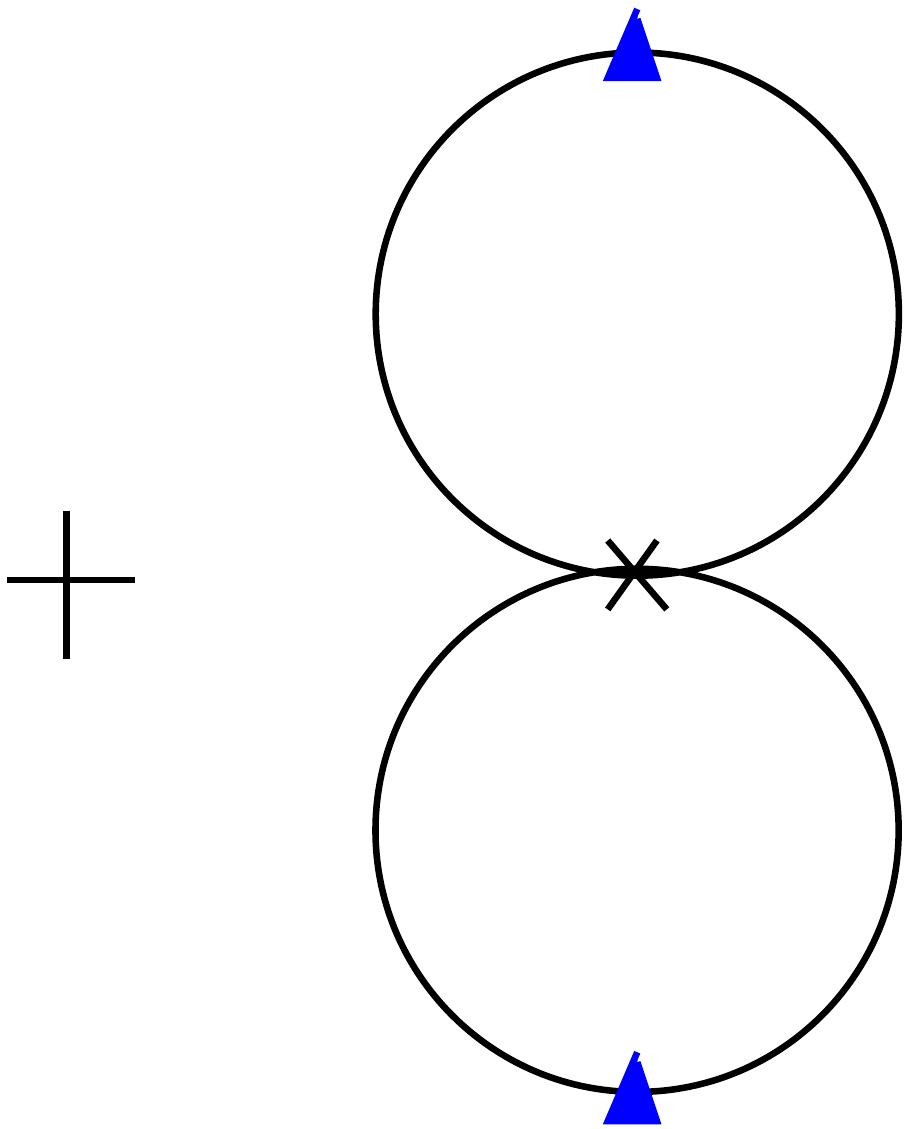}
	\caption{The scale anomaly gets contributions from the $d=4$ operators: the blob denotes the insertion of ${(F^a_{\mu\nu})}^2$ and the cross denotes the marginal four-Fermi operators, generated at scales below $\mu_{\rm cr}$. The solid lines are full propagators of techni-quarks and the curly lines are the full propagators of techni-gluons. }
	\label{fig4}
\end{figure}
Since the scale symmetry is spontaneously broken by the condensation of techni-quarks, 
a dilaton arises as a pseudo Nambu-Goldstone boson by Goldstone theorem:
\begin{equation}
\left<0\right|D^{\mu}\left|\sigma\right>=iF_{\sigma}p^{\mu}e^{-ip\cdot x}\,,
\end{equation} 
where $F_{\sigma}$ is the dilaton decay constant. If the dilaton pole dominates at low energy, we may take 
$\partial_{\mu}D^{\mu}=F_{\sigma}m_{\sigma}^2\,\sigma$ to get 
\begin{equation}
\left<\partial_{\mu}D^{\mu}\right>=F_{\sigma}^2m_{\sigma}^2\simeq \kappa^2\,m_F^4\,,
\end{equation}
where $\kappa$ is a constant of order one.
Therefore the dilaton can be quite light, $m_{\sigma}\simeq\kappa\,m_F^2/F_{\sigma}\ll m_F$, if $F_{\sigma}\gg m_F$.
Whether the dilaton decay constant can be much bigger than $m_F$ or not in WTC is not settled yet. But, I give a compelling argument that it is so. Let's assume that the chiral symmetry is not spontaneously broken but explicitly by a small mass, $m_0$. 
Then by the scale invariance the dilaton has to couple to the techni-quarks as 
\begin{equation}
m_0\,e^{\sigma/F_{\sigma}}\bar QQ\,.
\end{equation}
If we integrate out the massive techni-quarks, we get an effective potential
\begin{equation}
V(\sigma)=4\eta\,m_0^4\left(\frac{\sigma}{F_{\sigma}}\right)^4\left[\ln\left(\frac{\sigma}{F_{\sigma}}\right)-\frac14\right]\,,
\end{equation}
which gives mass to the dilation 
\begin{equation}
m_{\sigma}^2=\eta\,\frac{m_0^4}{F_{\sigma}^2}\,\to0,\quad {\rm if}~~m_0\to0\,.
\end{equation}
Since $\beta(\alpha)\approx0$,   the theory is scale-invariant even at quantum level, when the chiral condensation vanishes.  The potential for the dilaton is therefore flat and the vacuum has a flat direction along the scale transformation. The dilaton decay constant, which is nothing but the vacuum expectation value of the dilaton fields, can take an arbitrary value. By taking $m_0$ to be parametrically much smaller than $F_{\sigma}$ the dilaton can be made arbitrarily light.  Therefore, even in the case the chiral symmetry is spontaneously broken, generating the dynamical mass for the techni-quarks, the dilaton can be made light, since the dilaton develops its vacuum expectation value at scale higher than $\mu_c$ as long as the theory is scale-invariant
or $\beta(\alpha)\approx0$\,.

\section{Composite Higgs and light techni-dilaton}
Consider a following composite 
operator
\begin{equation}
\lim_{y\to x}Q(x)\bar Q(y)=\left(\mu\left|x-y\right|\right)^{\gamma_{m}}Q\bar Q(x)\,,
\end{equation}
where $Q$ is the techni-quark field and $\gamma_m$ is the anomalous dimension of the techni-quark bilinear operator. 
Since the electroweak symmetry is broken by the condensation of techni-quarks, we may write the composite operator as 
\begin{equation}
Q\bar Q(x)\sim \,H(x)=e^{i\pi_{\rm TC}/F_{\rm TC}}
\begin{pmatrix}
0  \\
v+h(x) 
\end{pmatrix} \,, 
\end{equation}
where  $v\equiv \left<H(x)\right>=247~{\rm GeV}/\sqrt{N_f}$ is the vacuum expectation value of the composite field $H(x)$, which transforms like a bi-fundamental, $\left(N_{f},N_{f}\right)$ under the chiral symmetry of WTC. The techni-pions 
$\pi_{\rm TC}$ are the excitations along the vacuum degeneracy and three of them constitute the longitudinal components of the 
electroweak gauge bosons, $W^{\pm}$ and $Z^0$. The radial excitation $h(x)$, orthogonal to the techni-pion excitations, is the composite Higgs, which is similar to the $\sigma$ or $f_0(500)$ isospin-singlet scalar meson. 

In QCD the $f_0(500)$ meson is not so light and is quite broad. However, unlike the technicolor of scaled-up QCD  the composite Higgs could be quite light in walking technicolor. For
WTC with techni-quarks in a two-index higher-dimensional representation, the composite Higgs is shown to be quite light, compared to the technicolor scale ($\sim1~{\rm TeV})$ in the large $N_{\rm TC}$ limit~\cite{Hong:2004td}.  Similarly, Kutasov 
{\it et al.}~\cite{Kutasov:2011fr} found in a holographic model of WTC the composite Higgs is quite lighter than techni-vector mesons in the conformal limit 
\begin{equation}
\frac{m_h}{m_V}\approx 0.2\,.
\end{equation}

Though composite Higgs and techni-dilaton have same quantum numbers at low-energy, they are created by two different 
composite operators, namely by dilatation currents and the techni-quark bilinear, respectively and transform differently under the scale transformation, $x^{\mu}\longrightarrow e^{-\lambda}x^{\mu}$, as following:
\begin{equation}
\sigma(x)\longrightarrow \sigma(x)+\lambda,\quad {\rm while}\quad h(x)\longrightarrow\,e^{\lambda}\,h(x)\,.
\end{equation}
Since they have same quantum numbers (except under the scale transformation), they do mix. The mixing can be read off from the effective Lagrangian for the composite Higgs:
\begin{equation}
{\cal L}_H=\frac12\left|D_{\mu}H\right|^2-\frac12 m_H^2\,e^{2\sigma/F_{\sigma}}H^{\dagger}H+\frac{1}{4}\lambda\,\left(H^{\dagger}H\right)^2+\,
\cdots\,.
\end{equation}
The mixing angle is therefore given as for a large $F_{\sigma}$
\begin{equation}
\theta\sim \frac{m_h}{F_{\sigma}}\,.
\end{equation}

By examining the recent data at LHC8, Matsuzaki and Yamawaki showed that $v/F_{\sigma}\simeq 0.2$ gives better $\chi^2$ fit than the SM, if one disregards the channels with lower significances~\cite{Matsuzaki:2012mk}. In this case the mixing is sizable and we have two light scalars at low energy. The newly discovered boson might be then a linear combination of both composite Higgs and techni-dilaton, which naturally explain both the enhancement in two-photon decay and the Tevatron data on 
$\bar b\, b$.

\section{Conclusion}
With discovery of new boson at LHC particle physics has now entered a new era. Though the current data is consistent with the 
SM, we believe there should be new physics beyond the SM. But, the current data indicates that the scale of new physics BSM might be much higher than the scale, proved at LHC8. 

One of the good candidates for BSM is walking technicolor, which breaks the electroweak symmetry dynamically. WTC predicts a light techni-dilaton if it is close to the conformal window. In addition to the techni-dilaton WTC predicts a light and narrow composite Higgs, which mix with the techni-dilaton. Therefore the newly discovered boson might be a mixture of these two light scalars. The phomenology of WTC with two light scalars is quite interesting and is being investigated. 

\section*{Acknowledgments}
We would like to thank the organizers, especially Koichi Yamawaki, for the invitation to SCGT12 and the hospitality during the conference. 
This work was supported by the Korea Research 
Foundation Grant funded by the Korean Government
KRF-2008-341-C00008\,.

\bibliographystyle{ws-procs975x65}
\bibliography{ws-pro-sample}

\end{document}